# Superconductivity beyond Pauli's limit in bulk NbS$_2$: Evidence for the Fulde-Ferrell-Larkin-Ovchinnikov state

Chang-woo Cho[1], Jian Lyu[1,2], Cheuk Yin Ng[1], James Jun He[3], Tarob A. Abdel-Baset[4,5], Mahmoud Abdel-Hafiez[6] and Rolf Lortz[1,*]

[1]Department of Physics, The Hong Kong University of Science and Technology, Clear Water Bay, Kowloon, Hong Kong.

[2]Department of Physics, Southern University of Science and Technology, 1088 Xueyuan Road, Nanshan District, Shenzhen, Guangdong Province, China.

[3]RIKEN Center for Emergent Matter Science (CEMS), Saitama, Wako 351-0198, Japan.

[4]Department of Physics, Faculty of Science, Taibah University, Yanbu, 46423, Saudi Arabia;

[5]Department of Physics, Faculty of Science, Fayoum University, Fayoum 63514, Egypt;

[6]Department of Physics and Astronomy, Uppsala University, Uppsala SE-75120, Sweden.

We present magnetic torque, specific heat and thermal expansion measurements combined with a piezo rotary positioner of the bulk transition metal dichalcogenide (TMD) superconductor NbS$_2$ in high magnetic fields applied strictly parallel to its layer structure. The upper critical field of superconducting TMDs in the 2D form is known to be dramatically enhanced by a special form of Ising spin orbit coupling. This Ising superconductivity is very robust against the Pauli limit for superconductivity. We find that superconductivity beyond the Pauli limit still exists in bulk single crystals of NbS$_2$. However, the comparison of our upper critical field transition line with numerical simulations rather points to the development of a Fulde-Ferrell-Larkin-Ovchinnikov state above the Pauli limit as a cause. This is also consistent with the observation of a magnetic field driven phase transition in the thermodynamic quantities within the superconducting state near the Pauli limit.

Transition metal dichalcogenides have been the focus of recent research due to their wide range of unique electronic properties in their 2D form with high potential for technological applications [1-8]. Among them are intrinsic superconductors with 2H-NbSe$_2$ and 2H-NbS$_2$ as representatives of the highest critical temperatures. In their 2D form, they have aroused great interest due to the discovery of Ising superconductivity, which allows them to withstand the Pauli limit of superconductivity [4,5]. While the upper critical field ($H_{c2}$) of a spin-singlet type-II superconductor is normally determined by the orbital limit for superconductivity [9], there are rare cases where the orbital limit is particularly high. Possible reasons are heavy effective masses of the quasiparticles [10-12], or a highly anisotropic structure that suppresses the orbital limit due to the open nature of the Fermi surface [13-24]. In this case, superconductivity is in principle abruptly destroyed at the Pauli limit [25,26] in the form of a first-order phase transition. There are two ways in which a superconductor can maintain its superconducting (SC) state above the Pauli limit. The first possibility is the formation of a Fulde-Ferrell-Larkin-Ovchinnikov (FFLO) state [27,28], in which the Cooper pairs obtain a finite center-of-mass momentum, resulting in a spatially modulated order parameter in a wide range of their magnetic-field vs. temperature phase diagram, which can extend far beyond the Pauli limit. Such spatial modulation of superconductivity has been described as a pair density wave state and is also used to explain the pseudogap phase in cuprates [29]. Prominent examples of the FFLO state have been found in some layered organic superconductors [13-23], but also in form of the prominent Q-phase the heavy-fermion compound $CeCoIn_5$, where a spatially modulated order parameter coexists with a magnetic spin density wave order [10-12], and most recently in the iron-based superconductor $KFe_2As_2$ [24]. For 2D dichalcogenides, Ising superconductivity is another possibility [4,5]. Here, the breaking of the mirror symmetry in the plane leads to a very strong pinning of the electron spins out of the plane due to the Ising-Spin-Orbit interaction (ISOI), resulting in opposite spin directions in the adjacent K and K' electron pockets. This ISOI effectively protects the Cooper pairs when an in-plane magnetic field is applied to the 2D layer, resulting in enormous enhancements of the critical field. This phase is of particular interest due to the recent theoretical prediction of a topological SC state with Majorana zero modes in high parallel applied fields in monolayer $NbSe_2$ [30,31].

---

[*] Corresponding author: lortz@ust.hk

While reports on Ising superconductivity usually focus on monolayers, it is also clear that $H_{c2}$ in 2D materials with multiple layers still exceeds the Pauli limit [5]. It is not clear whether ISOI could still have an influence in the bulk form of these layered materials with very weak interlayer coupling. In this letter, we present magnetic torque, specific heat and thermal expansion measurements on a bulk single crystal of $NbS_2$, where we align the layered structure parallel to the applied magnetic field using a piezo rotary stage with millidegree accuracy. We find that $H_{c2}$ significantly exceeds the Pauli limit of 10 T at low temperatures and shows a pronounced characteristic upswing towards low temperatures. This is a clear indication that an unusual SC state is formed [14,15,24,32]. Thermal expansion measurements as a bulk thermodynamic method, which are closely related to the specific heat, indicate a small transition anomaly near the Pauli limit indicating a phase transition within the SC state. Using theoretical simulations [33], we show that Ising superconductivity is not accountable for such a strong $H_{c2}$ upturn in bulk $NbS_2$. The existence of the additional phase transition within the SC state points to a formation of the FFLO state.

Figure 1a shows magnetic torque data measured at various fixed temperatures with a small 1° misalignment of the field with respect to the basal plane of $NbS_2$. At lower temperatures, when $H_{c2}$ reaches higher fields, it is obvious that the transition becomes sharper and more step-like, which may be an indication for Pauli-limited first-order behavior [15,24]. Only a small tail persists above the step-shaped transition, which indicates a certain persistence of superconductivity. $H_{c2}$ reaches a maximum at 1.15 K and then decreases again slightly towards 0.35 K.

To achieve a strictly parallel field orientation, we aligned the layered structure of the $NbS_2$ single crystal by first minimizing $\tau$ at a fixed field and temperature by gradually rotating the sample through the parallel orientation [24]. This allowed us to determine the approximate parallel orientation. Subsequently, we repeated field scans at tiny variations of orientation of 0.1° or less until we found torque data with minimum amplitude and minimal opening of the hysteresis loop. This corresponded to the parallel orientation. A sequence of measurements at small angular variation is shown in Fig. 1b. As the field decreases, the torque builds up much more continuously below $H_{c2}$, suggesting that the sharp jumps in Fig. 1a are due to screening currents that build up continuously during the field sweep in the SC state and decay abruptly when approaching the Pauli limit in slightly tilted fields.

In Fig. 2a, we show $\tau$ data measured at different fixed temperatures during field sweeps for the field applied parallel to the NbS$_2$ basal plane. Here we identify the $H_{c2}$ transition from the onset point where the data starts to deviate from zero, as marked by the additional open circles. In the same field, a hysteretic opening of the two branches, which was recorded when the field was swept up and down, confirms the onset of superconductivity. At 350 mK $H_{c2}$ exceeds our highest field of 15 T, and a hysteresis exists up to 15 T (Fig. 2b).

In Fig. 2c we present torque data measured in constant, strictly parallel fields as a function of temperature. The SC transition can be identified as a step-like transition. The data were normalized by the jump size at $T_c(H)$ for better clarity. We take the upper onset of the transition to identify $T_c(H)$ as marked by the additional open circles. Note that other criteria to define of $T_c(H)$ or $H_{c2}(T)$ provide qualitatively similar phase diagrams.

In Fig. 3 and Fig. 4 we summarize our specific heat and linear thermal expansion data. To achieve the parallel field orientation, we maximized the $T_c$ in a fixed field of a few Tesla by slightly rotating the sample in repeated measurements [24]. Fig 3a illustrates the specific heat $C/T$ and the linear thermal expansion coefficient $\alpha = (1/L_0)\, dL(T)/dT$ in a parallel field of 15 T. Both thermodynamic quantities, which are closely related via the thermodynamic Ehrenfest relationship, show a SC transition, which demonstrates that 15 T is not sufficient to suppress superconductivity. In the inset (Fig. 3b) we show specific heat data measured with our *ac* modulated temperature technique measured during field sweeps at constant temperature of the thermal bath. A standard BCS superconductor would show the characteristic step-like transition at $H_{c2}$ [34]. However, the NbS$_2$ data only display a broad bump centered at relatively low fields, while the signal fades very gradually towards higher fields to approach the normal state. There are no additional features to be seen here that could indicate additional phase transitions within the SC state. However, in Fig. 4a we illustrate the magnetic field derivative of the 300-mK specific heat data together with the linear magnetostriction coefficient $\lambda = (1/L_0)\, dL(H)/dH$. For both quantities, a small anomaly indicates a phase transition within the SC state at ~10 T, which occurs at the theoretical Pauli limit. The transition is reproducible when the field is swept up and down. In the following we will attribute it to the transition that separates the ordinary low-field SC state from a high-field FFLO state. In the magnetic torque $\tau(H)$ in the temperature range of 0.3 to 2.5 K tiny anomalies are hidden in the large slope near 10 T, but become visible after subtraction of a linear background in the form of a downward step, or as a dip-like structure in the field derivative $d\tau(H)/dH$ (Fig. 4b).

We summarize our results in the $H/T$ phase diagram in Fig. 5. At high temperatures, the $H_{c2}$ transition line begins to rise with a non-vertical slope. This is in contrast to the behavior found in 2D samples and shows that there are significant orbital effects on the upper critical transition, which are naturally absent in 2D samples. We also include scaled $H_{c2}$ data of the organic superconductor κ-(BEDT-TTF)$_2$Cu(NCS)$_2$ [15], for which numerous studies have demonstrated the existence of an FFLO state [13-15,18-21]. It is obvious that the initial slope of NbS$_2$ is much weaker than for κ-(BEDT-TTF)$_2$Cu(NCS)$_2$, which proves a smaller Maki parameter $\alpha_m = \sqrt{2}\frac{H_{\mathrm{orb}}(0)}{H_P(0)}$. A large Maki parameter in the range between 1.7-3.4 [35,36] is considered one of the decisive prerequisites for the formation of an FFLO state. From a fit with the Werthamer–

Helfand–Hohenberg (WHH) model of the initial $H_{c2}$ slope of NbS$_2$ we obtain an orbital limit $H_{orb}$ ~23 T, which gives $\alpha_m$ = 3.25, which is certainly large enough to support an FFLO state. At 3.6 K, the $H_{c2}$ line rises above the BCS Pauli limit at ~10 T and then begins to saturate down to 1.5 K. At lower temperatures it shows a characteristic upswing and raises above 15 T, where it leaves the field region accessible in our experiment. The upswing is strikingly similar to that observed in κ-(BEDT-TTF)$_2$Cu(NCS)$_2$, with both data being almost congruent in the high field region. This upswing is suppressed if the NbS$_2$ sample is misaligned by only one degree. Such an upswing indicates a certain change in the SC properties making it more robust against the strong Zeeman fields. It is generally interpreted as an indication of the development of an FFLO state. In fact, the phase diagram looks remarkably similar to other FFLO systems [10-24], including the field-induced transition at ~10 T, which probably indicates the phase change between the ordinary SC state at low fields and the high-field FFLO state. However, in most other FFLO systems, the $H_{c2}$ transition typically sharpens and becomes of first order as one approaches the Pauli limit [11,14,24]. In NbS$_2$, however, the $H_{c2}$ transition in high parallel fields remains very continuous, indicating a very gradual decay of superconductivity towards the normal state in the form of a second order transition. In Ref. 37 it was shown that the first-order nature of the $H_{c2}$ transition in high fields occurs only in the absence of the orbital effect, while in its presence it remains of second order anywhere. With orbital effect it is expected that only the transition between the FFLO state and the BCS state should be of first-order nature. This is fully consistent with our observation: The finite $H_{c2}$ slope in the high temperature range is clearly due to the orbital effect, and the additional transition at 10 T, observed as a spike in the magnetostriction coefficient, indicates a first-order nature of the FFLO to BCS transition. However, a small misalignment of only one degree sharpens the transition observed in torque to a step-like more first-order-like transition and indicates that superconductivity becomes Pauli limited. The fact that the $H_{c2}$ transition line then goes through a maximum is indeed expected for the Pauli limited case in tilted fields [37].

It is often claimed that ISOI is a specialty of monolayers with their broken inversion symmetry, which leads to a spin splitting of the Fermis surface, while it is spin degenerate in bilayers and the bulk [5]. In reality bilayers have $H_{c2}$ values, which still exceed the Pauli limit by a factor of ~4 [5]. This suggests that ISOI can be weakened by the intra-layer coupling but is not completely quenched because of weak intra-layer coupling in TMD materials. Therefore, one concern regarding the FFLO scenario as an explanation of the phase diagram is that the ISOI should still have a considerable influence in the bulk. ISOI is not compatible with a finite-momentum pair density wave state in the FFLO phase because it suppresses the effect of the in-plane Zeeman field on the Fermi surface. However, NbS$_2$ has multiple Fermi pockets at the Γ and K/K' points. While the K pockets are protected by ISOI, the spin-orbit interaction at the Γ pocket is weak and therefore compatible with the FFLO state. To test whether the phase diagram could alternatively be explained by Ising superconductivity, we performed numerical simulations using a three-band tight-binding model [38] with a spin-orbit coupling strength of $\beta_{SO}$ = 87meV. In order to simulate a bulk sample, we used a thickness of 10 layers, which proved to be sufficient since we found that increasing the thickness further does not affect the critical field much for greater thicknesses. The obtained low-temperature $H_{c2}$ line is included in Fig. 5. While for samples in the 2D limit it could be shown that Ising superconductivity can cause very pronounced $H_{c2}$ upturns at low temperature similar to FFLO states [6,33,39], it is obvious that for reasonable values of the spin-orbit coupling strength the upturn of bulk $NbS_2$ is much too weak to explain our phase diagram.

To conclude, with the additional evidence of a field-driven phase transition within the SC state near the Pauli limit and the good agreement with theoretical simulations, it can be concluded that the phase diagram we observed for a bulk $NbS_2$ single crystal in magnetic fields strictly parallel to its layer structure, with its pronounced upswing of the $H_{c2}$ line far above the Pauli limit for superconductivity, is most plausibly explained by an FFLO state that forms in magnetic fields above 10 T. This high-field phase requires further systematic experiments including nuclear magnetic resonance to directly monitor the nature of the FFLO pair density wave state.

**Acknowledgments**

We thank U. Lampe for technical support. This work was supported by grants from the Research Grants Council of the Hong Kong Special Administrative Region, China (GRF-16302018, GRF-16300717, C6025-19G, SBI17SC14).

**Author contributions**

This work was initiated by R.L.; C.w.C, J.L. and C.Y.N. carried out the magnetic torque experiments. C.Y.N., J.L. and R.L. carried out the specific heat experiments; the thermal expansion experiments were conducted by C.Y.N. and R.L.; the single crystal sample was provided by M.A.H. and T.A.A.B.; J.J.H. provided the numerical simulations. The manuscript was prepared by R.L. and all authors were involved in discussions and contributed to the manuscript.

**Competing financial interests**

The authors declare no competing financial interests.

**Data availability**

The data that support the findings of this study are available from the corresponding author upon reasonable request.

## Methods

### Sample Preparation

The high-quality single crystal of 2H-NbS$_2$ was grown by a solvent evaporation technique, which is described in detail in Ref. 40. The sample represented a small square shaped platelet, which was completely flat on the macroscopic scale. The zero-field specific heat (Fig. 6) displays a sharp superconducting transition jump centered at $T_c$ = 5.5 K indicative for a large superconducting volume fraction and good quality of the single crystal.

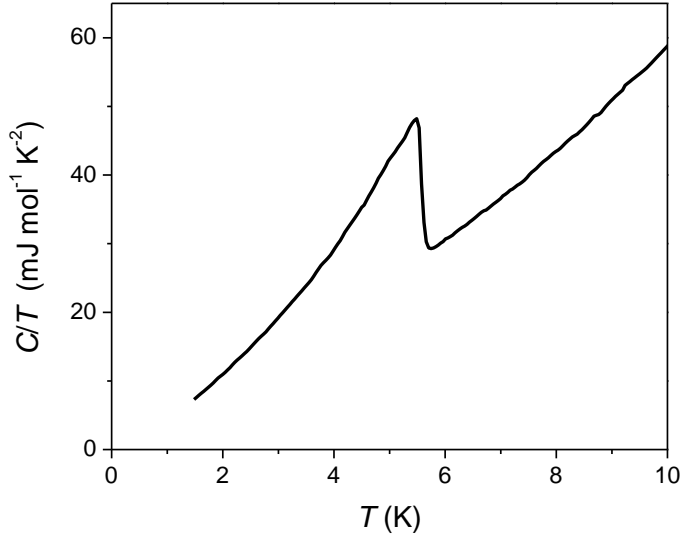

FIG. 6. Zero-field specific heat $C/T$ of the NbH$_2$ single crystal used in this study showing a sharp superconducting transition centered at $T_c$=5.5 K.

### Experimental techniques

All our experimental sensors fit on the same piezo rotary stage mounted on our $^3$He probe in a 15 T magnet cryostat. The rotator allows the sample to be aligned in relation to the field direction with millidegree accuracy. Field sweeps were conducted at a rate of 0.1 - 0.5 T/min. Temperature dependent measurements at different fixed fields were performed during sweeps at 0.04 K/min.

The magnetic torque $\tau$ was measured using a capacitive cantilever technique [24]. The cantilever leg is insulated from the counterplate of the capacitor by a thin sapphire sheet. This allows reversible measurements of the torque both as a function of the field or temperature. The capacitance is measured with a General Radio 1615-A capacitance bridge in combination with a SR850 lock-in amplifier. The torque $\tau$ is directly related to the anisotropic DC magnetization by the relation $\tau = \mathbf{M} \times \mathbf{H}$, where $\mathbf{H}$ is the applied magnetic field. Since for such layered superconductors the DC magnetization is expected to be greatest in the out-of-plane direction, this relationship suggests that $\tau$ vanishes in parallel fields. In reality, it reaches a minimum, but does not disappear completely during a complete field sweep due to higher-order quadrupole components [15,24]. The ultra-high resolution allows us to detect the tiny signal with a very good signal-to-noise ratio.

The thermal expansion was measured with a miniature capacitance dilatometer [41], in which the sample is pressed with a screw mechanism against one of the plates of a parallel-plate capacitor suspended by a firm spring mechanism. A change in the sample length, induced either by a change in temperature (thermal expansion) or by the field (magnetostriction), changes the distance between the plates and thus causes small changes in the capacitance, which is measured in the same way as described for the torque. Thermal expansion is a bulk thermodynamic quantity closely related to specific heat and allows us to detect small changes in the sample length that occur during phase transitions.

The specific heat $C_p$ was measured using an alternating temperature (*ac*) technique with a mK temperature modulation amplitude [24]. To account for the relatively flat temperature dependence of the $H_{c2}$ line in the low temperature / high magnetic field (*H/T*) regime, we performed the experiments at fixed base temperature during field sweeps. Note that the data measured in this somewhat unusual way still represent the thermal response of the sample with respect to a small temperature change, although here we probe the variation of the specific heat with respect to the field. The reason for is that near the Pauli limit the slope of the $H_{c2}$ line in the magnetic field vs. temperature (*H-T*) phase diagram becomes very small and measurements during field sweep sampling reveal much more details in the important part of the phase diagram. The calorimeter platform was supported by thin nylon wires that become stiff at low temperature and serve as a support to prevent magnetic torque effects.

**Figures**

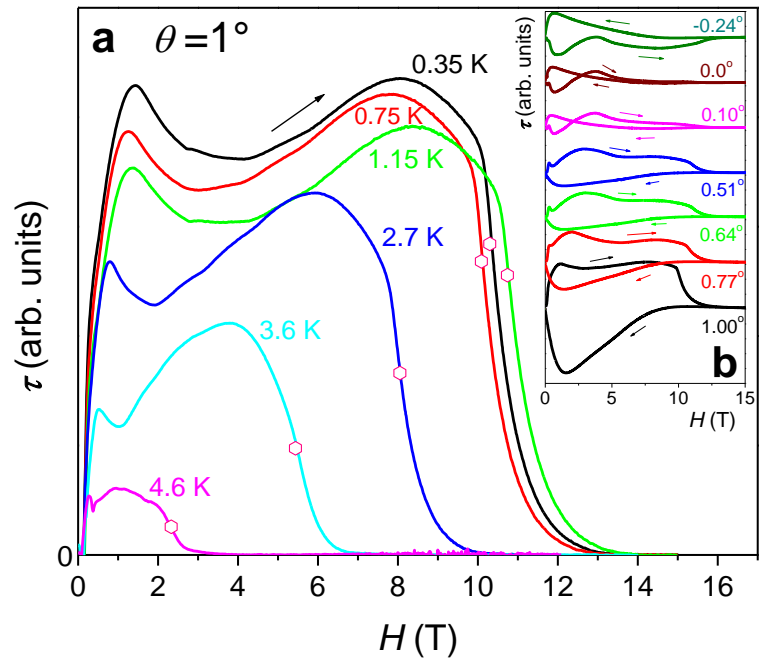

**Figure 1 | Magnetic torque data in a slightly tilted field of NbS$_2$. a** Magnetic torque $\tau(H)$ of NbS$_2$, measured at fixed temperatures as a function of the applied magnetic field $H$, oriented with a small angle $\theta = 1°$ to the basal plane. The data were measured at increasing field. Circles mark the points where the steepest slope occurs near $H_{c2}$. **b** Magnetic torque measured at $T=0.35$ K at various angles near $\theta = 0°$ (parallel field). A weak linear normal state contribution was removed from all data and offsets were added for better clarity.

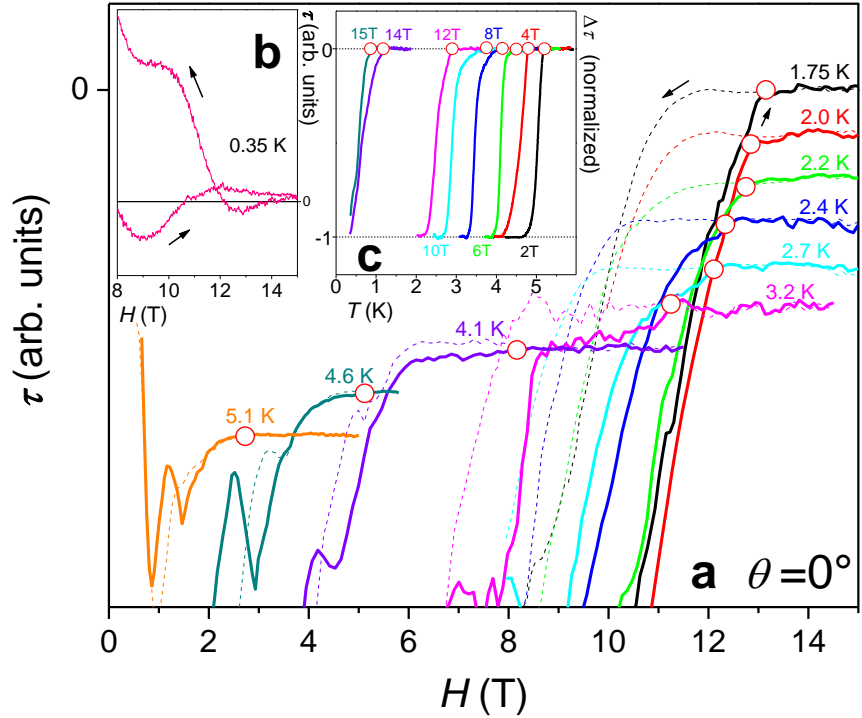

**Figure 2 | Magnetic torque data in parallel fields of NbS$_2$. a** Magnetic torque measured at fixed temperatures as a function of the magnetic field applied strictly parallel to the basal plane ($\theta = 0°$) near the onset upper critical field $H_{c2}$ (open circles). The solid lines represent data measured at increasing field, while the dotted lines are data measured at decreasing field (overlapping parts of the data for low fields have been omitted and offsets have been added for reasons of clarity). **b** Magnetic torque data at 0.35 K where $H_{c2}$ exceeded the highest field in our magnet cryostat. **c** Magnetic torque measured in fixed parallel fields as a function of temperature. The data was normalized and a weak linear normal state background was removed for clarity. Open circles mark the onset critical temperatures.

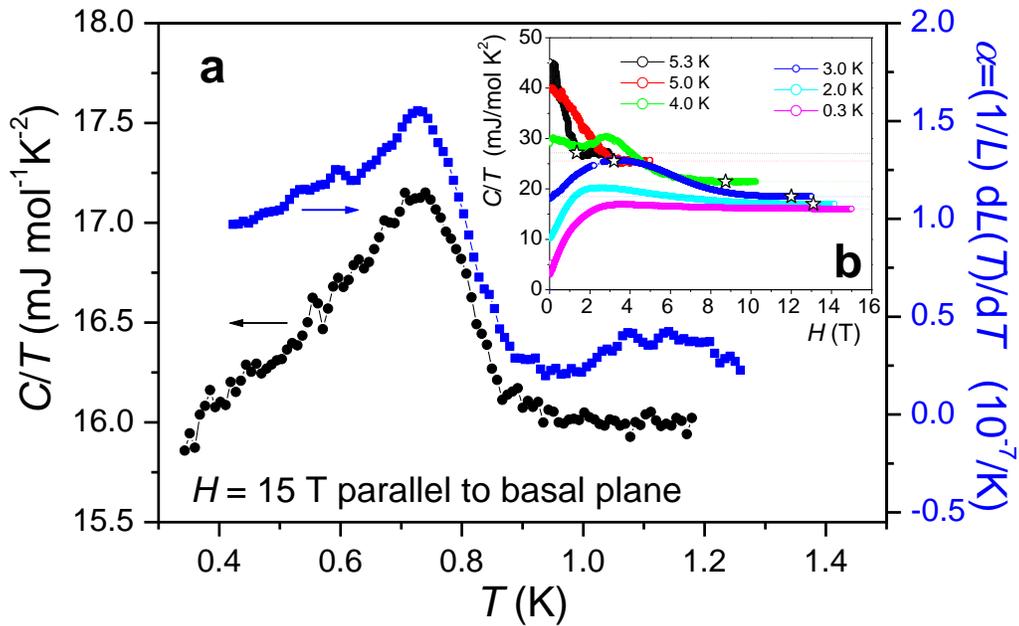

**Figure 3 | Thermodynamic bulk probes displaying the upper critical field $H_{c2}$ transition of NbS$_2$ in high parallel magnetic fields. a** Specific heat $C/T$ (circles) and linear thermal expansion coefficient $\alpha = (1/L_0)\, dL(T)/dT$ (squares) of NbS$_2$ showing the superconducting transition in a magnetic field of 15 T applied strictly parallel to the layer structure. **b** Specific heat of NbS$_2$ measured with an *ac* technique with a small 1 mK temperature modulation during field sweeps at different fixed temperatures. The stars mark the fields in which the constant normal state specific heat value (dotted lines) is reached.

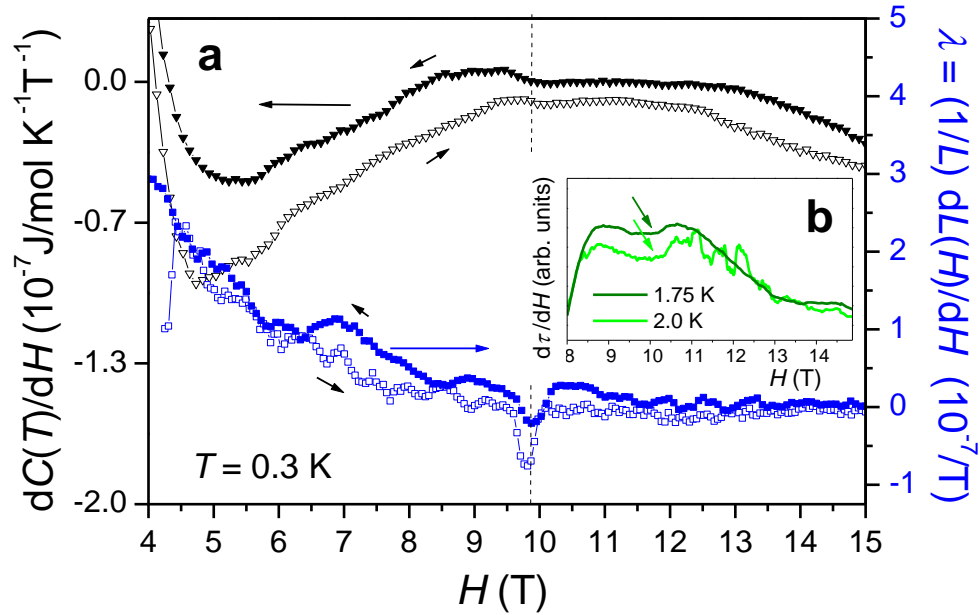

**Figure 4 | Thermodynamic bulk probes displaying the phase transition between the BCS low-field SC phase and the FLLO state in NbS$_2$. a** Magnetic field derivative $dC(T)/dH$ of the specific heat, measured at 300 mK during a field sweep (see **b** for the corresponding $C/T$ data) together with the linear magnetostriction coefficient $\lambda = (1/L_0)\, dL(H)/dH$ for magnetic fields applied strictly parallel to the layer structure. Both quantities reproducibly show a small transition anomaly at ~10 T. **b** Field derivative $d\tau(H)/dH$ of the magnetic torque for selected temperatures showing a small dip-like anomaly near 10 T.

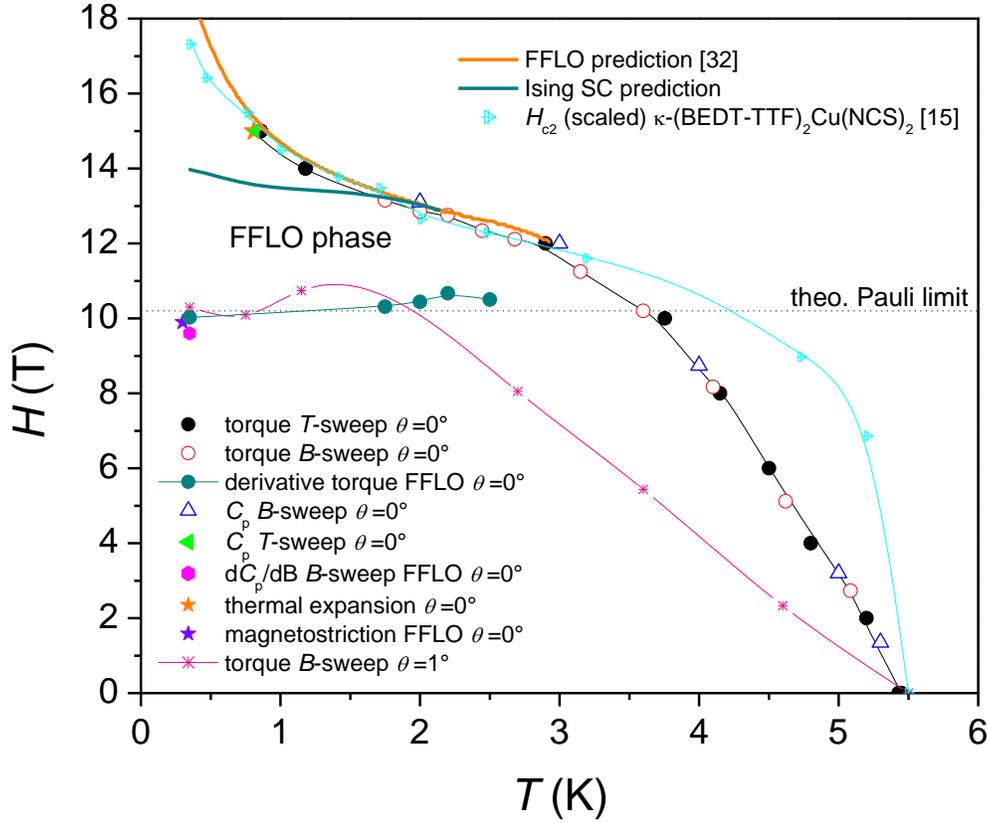

**Figure 5 | Magnetic field vs. temperature phase diagram of NbS$_2$ in magnetic field applied parallel to the layer structure.** Filled black circles represent the upper critical field $H_{c2}(T)$ as measured by torque magnetometry during field scans at constant temperature, red open circles represent the critical temperature $T_c(H)$ measured by torque during temperature scans. Blue triangles are $H_{c2}(T)$ data obtained from ac calorimetry measured during field scans. The orange star and green filled circle mark the critical temperature in $C_p(T)$ and $\alpha(T)$ data measured in 15 T. The violet star marks a small additional transition anomaly in the magnetostriction and filled magenta circle marks a similar anomaly visible in the field derivative of the specific heat near the Pauli limit due to the transition to the FFLO state (Fig. 2c). In addition, the dark cyan circles correspond to a small anomaly in the magnetic torque attributed to the same phase transition within the superconducting state (Fig. 2d). Additional lines mark $H_{c2}$ data obtained from torque magnetometry in field scans measured with a small misalignment angle of 1 degree. The orange and dark cyan lines are theoretical predictions of the low temperature $H_{c2}$ line for superconductivity with the FFLO state [32] and for Ising superconductivity, respectively. We have also added a scaled $H_{c2}$ curve of the organic superconductor κ-(BEDT-TTF)$_2$Cu(NCS)$_2$, which shows a very similar upturn of the $H_{c2}$ line due to the formation of an FFLO state [15].